# Analysis of Personal Data Visualization Reviews On mHealth Apps

**Yasmeen Anjeer Alshehhi**
yanjeeralshehh@deakin.edu.au
Deakin University

**Mohamed Abdelrazek**
Mohamed.abdelrazek@deakin.edu.au
Deakin University

**Alessio Bonti**
a.bonti@deakin.edu.au
Deakin University

## Abstract

Mobile devices, specifically smartphones, provided easy and quick access to data visualisations throughout various tracking apps. Mobile health (mHealth) apps have given non-expert users access to data visualisation to track their activities and health-related issues such as heart tracking and medication. However, no work is done on user experience or perception of data visualisations in mHealth apps. App reviews offer an indirect anchor for researchers to examine how non-expert users perceive and interact with data visualisations and identify the key challenges and recommendations.

This paper introduces an analysis of app reviews on data visualisations reported on a dataset of 250 mHealth apps on the Google Play Store. We identified 8,406 comments related to data visualisations. 919 neutral comments, 1,557 negative comments and 5,930 positive comments. From analysing the user reviews, functional requirements turned out to be the most common problem across these app reviews, followed by the look and feel and then data problems. A complete set of data visualisations seem to be the most well-received capability of mHealth apps. We used these comments to develop classification and data visualisation guidelines when developing mobile data visualisations.

## Introduction

The easy access to mobile devices made data visualisations widely adopted by non-experts for several personal needs. Tracking health data is one of the most common personal tracking aspects (Mamykina, et al., 2017). End-users used data visualization as a communication tool (Archambault, Helouvry, Strohl, & Williams, 2015), decision-making (Aparicio & Costa, 2015) and goals tracking (Choe, Dachselt, Isenberg, & Lee, 2019).

Personal data visualisations rely on mobile devices as the primary platform for user interactions (Lee, Choe, Isenberg, Marriott, & Stasko, 2020). However, mobile data visualisations introduce new display size, resolution, computing power, storage, and interaction modality challenges. Consequently, the field of visualisation interface design needs improvement to achieve users' satisfaction and cope with the challenges of the new environment (Gu, Mackin, & Zheng, 2018). Meanwhile, the industry has been active in proposing its own researched designs models – e.g. Google and IBM have created their design language incorporating data visualisation and design principles (Data Visualization , material design). However, these frameworks are limited to design principles only. The most common guidelines are desktop visualisation, with different environments and interaction methods from mobile devices. Although research studies have been comprehensive and continuously evolving to provide better user experiences for mobile data visualisation, users still report issues and challenges with their data (Gu, Mackin, & Zheng, 2018) , (Katz, Dalton, Holland, O'kane, & Price, 2017).

Graphs, charts, and icons are the main components of health tracking apps that help users track their goals, habits, and achievements (Choe, Dachselt, Isenberg, & Lee, 2019). Commendable efforts have been carried out in data visualisation related to diabetes, but evident gaps in designing and understanding data visualised in tracking apps are still present (Choe, Dachselt, Isenberg, & Lee, 2019. These gaps are related to the diversity of data visualisation audiences (Jena, et al., 2021) and the lack of generic guidelines that suit non-expert users and mobile devices (smartphones). Thus, there is a need to examine users' perspectives towards data visualisation in mobile apps and understand the context of the data visualisation usage to achieve better data visualisation on mobile devices Choe, Dachselt, Isenberg, & Lee, 2019, (Huang, et al., 2014). In this paper, we address the following questions:

- RQ1: What are the common visualisation tasks and charts that have been adopted in mHealth apps?
- RQ2: What are the top data visualisation issues in health tracking apps?
- RQ3: What guidelines can we provide based on user reviews to improve user experience?

User feedback is currently applied to seek users opinions and satisfaction of a specific service, or product (Hoon, Vasa, Schneider, Grundy, & others, 2013). For example, the industrial sector used this communication method to evaluate the quality of products and services based on user experiences to improve their services. The same concept applied to apps reviews in which users give either positive or negative comments (Hoon, Vasa, Schneider, Grundy, & others, 2013). To understand how these reviews are used, in 2008, Gebauer et al. demonstrated that user reviews assist in delivering software quality features and user requirements (Gebauer, Tang, & Baimai, 2008). In addition, it helps developers to iterate their apps to achieve user satisfaction and solve problems reported in the comments.

Further, these comments assist other end users in deciding on app downloading or purchasing (Khalid, 2014). Mobile health apps are used by multi-individuals from different nations, educational backgrounds, races, and ages. User reviews are crucial to exploring user satisfaction and investigating the missing quality attributes and users requirements. In 2018, Caldeira et al. published a study of 32 mood tracking apps with an analysis of 1,000 reviews. Their primary finding was that used data visualisation needed to be varied to match multiple people preferences (Caldeira, et al., 2017).

To the best of our knowledge, there is no work published in the literature that reviewed and analysed user app reviews and comments regarding data visualisation components of the app. Thus, in this paper, we introduce a detailed review of app reviews on a sample of 250 mHealth apps as the primary focus of our research. We believe that many of our findings can be expected (generalised) to other app domains. We are planning to extend our study to consider other areas and also view reviews on different app stores, including Apple App Store.

## Data collection

We queried the Google Play Store to collect health tracking apps and related information. Google Play restricts results to 250 apps, so we ran it on tracker and tracking and got 500 apps with some duplicates that we removed and ended up with 217 unique mobile health applications (mHealth) apps.

We had 2,750,000 app reviews for the 217 apps. The comments are stored in a table with four columns: AppID, Score, URL and Text. After removing duplicate and undefined rows, the remaining rows have dropped to 10,601. We filtered the reviews associated with visualisation terms (graph, chart, visual) and ended with 8,406 rows (reviews) related to data visualisation. The list of the filtered comments can be found in this anonymous link \href{https://www.dropbox.com/sh/sfmnqj9309dxgbj/AAC7YG-7Wi5BpIksNmgKs-1Da?dl=0}{(Dropbox link of the app reviews)}. Table 6 shows an example of generated comments.

# Results and analysis

This section addresses our research questions introduced in Section 1 using data collected in Section 2. We first present an initial analysis of the user reviews vs review sentiment, then analyse and reflect on the research questions.

## User reviews sentiment

We found that star-based rating does not accurately represent their concerns in their app review -- i.e. some comments had scored high ratings. However, still, it contains negative thoughts related to the data visualisation aspect.

Instead, we will indicate the positive and negative reviews based on their occurrences in each group as per our manual review. Table 1 shows the number of comments in each star rating. For example, in 4 stars rating, the number of negative reviews outweigh the number of positive reviews (59 \% negative reviews). Also, as can be seen, more than 7,000 out of 8,406 reviews are rating these apps as 4 or 5 stars. Yet we found more than 900 negative comments and more than 700 neutral comments out of these 7,000 comments.

*Table 1 App reviews: star rating vs review sentiment*

| Rating | Positive | Negative | Neutral | Total |
| --- | --- | --- | --- | --- |
| 1 & 2 | 13 | 283 | 138 | 434 |
| 3 | 51 | 302 | 143 | 496 |
| 4 | 366 | 591 | 276 | 1233 |
| 5 | 5500 | 381 | 362 | 6244 |
| Total | 5930 | 1557 | 919 | 8406 |

## RQ (1) What are the common visualisation tasks and charts that have been adopted in these mHealth apps?

As table 2 shows, We found 12 charts adopted in our dataset of 217 apps: Waterfall, stock, table, timeline, scatter, map, line chart, calendar, area chart, bar chart, pie chart, maps and Iconography.

Based on the data collected, Iconography was the most adopted visualisation type (141) as it was used in indicating statues using numbers or texts with related symbols and colours. Line and bar charts are the second most common charts (83) to track. Stock, waterfall, and scatter plot occurrence were the least (1).

Fourteen apps do not include any charts, in which users commented that they needed a chart to track their progress and goals. 4 apps have the maximum number of graphs (5) (Anxiety Tracker - Stress and Anxiety Log Blood Pressure Diary, Heart Rate Monitor, morePro), three are rated with more than four stars and one app rated with 2.5.

Maps occurred 21 times as a visualisation type due to the users tracking their walking, running, and cycling habits. These maps include information such as numbers of steps, starting and endpoint. The calendar is mainly used to track mood, period, pregnancy, and pills are taken (32). Five charts are adopted in the top best apps: line chart, area chart, pie chart, calendar, and bar chart. Other visualisation types adopted are colours and Iconography to indicate highs and lows moods, respectively, which have been used for monthly or yearly tracking, creating patterns, and seeing trends.

In terms of visualisation tasks, developers have focused only on one primary mission: tracking progress. These apps did not cover other visualisation tasks - e.g. comparing, filtering, ordering, grouping, maximum and minimum. Therefore, there is a lack of interaction with the presented charts.

*Table 2 Adopted chart types in health tracking apps*

| Charts | No. of occurrences |
|---|---|
| Stock | 1 |
| Waterfall | 1 |
| Table | 2 |
| Timeline | 4 |
| Scatter | 1 |
| Map | 21 |
| Line chart | 67 |
| Calendar | 32 |
| Area chart | 44 |
| Bar chart | 83 |
| Pie chart | 83 |

## RQ (2) What are the Top Data Visualisation Issues?

We manually reviewed the comments reported by app users to extract data visualisation issues and list them in an excel file. Next, we grouped the user reviews into 18 metrics (M1 to M18), as shown in Table 4. We further grouped these metrics into dimensions and criteria (discussed in Section 4).

Missing functional requirements were the most frequent negative reviews. For example, 50 \% of the applicable requirements issues are related to missing some needed graphs such as weekly or monthly charts, progress charts and charts to track aspects such as blood pressure, heart pulse, baby feeding. Other complaints were about the functionality and display of the graph. Additionally, there are four main issues related to data visualisation design (look and feel):

- Chart scaling issues (40 \%)
- Graph styling (i.e., font size, chart ranges, and colours do not indicate meaningful information, including grid lines in the charts and axis titles)
- Interactivity (scrolling, zooming, landscape, choose graphs options)
- The chart type is not appropriate (bar chart, pie chart and bar chart)

Data is another primary aspect that users criticised. It includes issues related to insufficient and incorrect data presented in the graph. In contrast, there are more general positive reviews than negative reviews that are related to users' satisfaction which include "love this graph" and" like this chart" without specifying any unique feature of the chart or graph (single and app visualisation).

The other aspect that users rarely mention is device capabilities and adaptability. The issues are summarised as follows:

- Being able to rotate the device to see the graph in landscape mode and visualisation consistency between the two operating systems, android and IOS
- Chart fitting with screen size

*Table 3 Summary of issues*

| Issues | Count | Percentage |
|---|---|---|
| Missing graphs – missing functionalities | 948 | 34% |
| Not the right charts | 22 | 0.7% |
| Charts are mixed up – charts displaying | 60 | 2% |
| Missing the chart type -font – no interactivity – colours – zooming problem | 224 | 8% |
| Chart scaling, graph layout -graph colors – small graph size | 404 | 14% |
| Zooming is not working well – resizing – graphs lines are mixed | 128 | 4% |
| Missing graph information | 202 | 7% |
| Not accurate info- charts units | 98 | 3% |
| Not showing information correctly | 32 | 11% |
| Missing the ability of phone rotating | 10 | 0.3% |
| Scale is not suiting screen size | 4 | 0.1% |
| Not fit with screen size - Different OS have different functionality | 36 | 1.2% |
| charts not working – two colour menus confusing | 154 | 5.5% |
| Visualization is meaningless | 74 | 2.6% |
| Low quality of graphs – charts are not always shown | 94 | 3.3% |
| Missing tooltips | 20 | 0.7% |
| Not good visual graph/chart | 10 | 0.3% |
| Charts are going to the right side of the screen – no consistency in showing graphs | 4 | 0.1% |

## RQ (2.1) Critical user concerns on mobile data visualisations in mHealth apps?

For this research question, we present statistics related to the number of users' reviews and show the ten best and worst apps based on the users' comments regarding the data visualisation. As shown in Table 5, these ten apps covered various health aspects. Furthermore, these apps are non-free apps rated with more than 4.5 stars. Therefore, we considered them good apps due to the number of positive reviews related to data visualisation compared with negative thoughts.

*Table 4 Best and worst apps*

| Best 7 apps | Worst 5 apps |
|---|---|
| Blood Glucose Tracker | Withings Health Mate |
| Daylio | Blood Glucose Tracker |
| mySugr | Baby Tracker |
| Baby Tracker | Weight Loss Tracker |
| Sleep as Android: Sleep cycle smart alarm | Period Tracker, Ovulation Calendar |
| Calorie Counter - MyNetDiary, | |
| Food DiaryTrack | |

*Table 5 Examples of positive and negative users' reviews related to functional requirement*

C1: A lot of info you can put on here. It would be nice to see multiple graphs for weight blood pressure. Even an option to send info as an attachment to emails for our doctors-- negative

C2: As a glucose tracking log, it is easy to use, and the chart is handy. I want a similar chart for A1C results. If it linked to my meter, it would be awesome --negative

C3: An excellent app. I wish it graphically displayed blood pressure the way it does blood sugar, but it's still the best blood sugar app I've found –harmful.

C4: Charts and graphs your blood sugar and tracks--positive

C5: Does just what I needed it to. It gives you the average and plots a graph to show your entries quickly. Thanks -- positive

C6: I love it brilliant at visually showing me how I've been and can predict my moods even bought the entire app – positive.

C7: Being newly diabetic this app has been beneficial in keeping things logged. It's also given me some comfort as I can see the effort I put in as straightforward to read results through their graphs and whatnot. I would recommend this app to anyone.--positive

C8: Has helped me visualise and control my blood glucose levels --positive

C9: The easy-to-read graphs show when I am in the red or green and ultimately turn my blood sugar down. --positive

C10: For the feeding graph, the colours blend too much and hard to see the difference. Another colour scheme would be favourable.--negative

**Functional requirement:** These apps are the highest-rated apps in terms of data visualisation among 217 apps. However, the comments in table 6 show that there are still some missing graphs that could be added, such as weight blood pressure(C1), A1C (a simple blood test that measures your average blood sugar levels over the past three months) (C2) \cite{website4}, blood sugar (C3) results. In contrast, users are satisfied with the other functions provided in the app related to data visualisation, such as charting entries (C4) and (C5).

**Data:** this dimension is related to the readability of the data. Users have commented that they are happy that they can obtain their progress through the charts: (C6), (C7) and (C8).

**Look and feel styling and interactivity** are the most important aspects that users' have mentioned in their positive reviews, such as C9 comments. However, some users were not satisfied with the colour selection (C10)

## Discussion

in this section, we discuss the implication of developing and designing data visualisation for mobile devices:

**Developing data visualisation**

We argue that data visualisation development needs special consideration. For example, adopting a complete set of functional requirements and data related to users control and understanding of the graphs. That is, users should control their entries and chart their progress easily. Further review associated with the functional needs is related to the device compatibility. For example, the touch, tap and rotating interaction are facilities provided by smartphone manufacturers that enable users to navigate and control their app screen easily.

**Designing data visualisation**

Different sectors, such as academic research and industries, have considered data visualisation design, as we mentioned in the introduction. However, still, there are some limitations related to smartphone designing. These limitations are chart size, scale, colour adoption, font size, data and interactivity.

**Including data visualisation in the app review**

Since app review does not include charts evaluation in the app feedback, it was not easy to find a protocol to classify the apps based on data visualisation reviews. So, we recommend adding data visualisation feedback in the app review as it is a central part of health tracking.

## Work Limitation

In this section, we explain the limitation of this paper. There are three main unexpected issues.

- Firstly, the apps considered were provided by an existing Google Play Store Search API where we can only offer keywords, including a list of apps. So we used (health tracking apps \& health tracker apps) keywords in our search. This API returns a list of apps whose titles match the search keywords according to Google.
- Secondly, the API is limited to only 250 apps. After running two queries around tracking and removing duplicates and unrelated apps, we ended up with 217. We plan to consider more questions and apps and more app categories in the future.
- Thirdly, the number of comments we got was not as expected, as there were many blank rows. Further filtering has reduced the number of words as there were no related comments even after using (charts, graphs, and visuals) to filter out the visualisation comments.

# Conclusion

We presented a manual analysis and categorisation of app reviews of data visualisations in 217 health tracking apps. We reported on these apps' adopted charts and tasks and evaluated 8,406 comments related to data visualisations. We found that the number of positive reviews is more than the negative reviews, so users are generally happy about the data visualisations in these apps. However, missing visualisations, look and feel and missing or incorrect data were the top concerns reported by mHealth app users. We found that collaboration between developing and designing data visualisation would enhance user experiences in health apps. We also argue that data visualisation feedback should be incorporated with app review. We plan to develop a data visualisation framework that cooperates with both developing and designing conduct a usability evaluation to investigate the effectiveness of the proposed framework in the future.